\documentclass[journal,comsoc]{IEEEtran}
\usepackage[T1]{fontenc}
\usepackage{times}
\usepackage{soul}
\usepackage{url}
\usepackage[hidelinks]{hyperref}
\usepackage[utf8]{inputenc}
\usepackage{graphicx}
\usepackage{amsmath}
\usepackage{amsthm}
\usepackage{booktabs}
\usepackage{algorithm}
\usepackage{algorithmic}
\usepackage{multirow}
\usepackage{threeparttable}
\usepackage{amsfonts}
\usepackage{float}
\usepackage{url}
\usepackage{bm}
\usepackage[table,xcdraw]{xcolor}
\usepackage{subfigure}
\usepackage{stfloats}
\usepackage[justification=centering, small]{caption}
\usepackage[numbers,sort&compress]{natbib}
\newcommand{\cmmnt}[1]{}
\usepackage{tabularray}
\usepackage{bbding}
\usepackage{amssymb}

\usepackage{pifont}
\newcommand{\cmark}{\ding{51}}
\newcommand{\xmark}{\ding{55}}

\ifCLASSINFOpdf
\else
\fi
\usepackage{amsmath}
\usepackage[cmintegrals]{newtxmath}
\begin{document}
%
\title{DQ-Data2vec: Decoupling Quantization for Multilingual Speech Recognition}
%
%
%


\author{Qijie Shao,
        Linhao Dong,
        Kun Wei,
        Sining Sun,
        and Lei Xie,~\IEEEmembership{Senior Member,~IEEE}

\thanks{Corresponding author: Lei Xie.}
  
\thanks{Qijie Shao, Kun Wei, Sining Sun and Lei Xie are with the Audio, Speech and Language Processing Group (ASLP), School of Computer Science and Engineering, Northwestern Polytechnical University, Xi’an 710072, China. Email: qjshao@npu-aslp.org (Qijie Shao), ethanwei@mail.nwpu.edu.cn (Kun Wei), snsun@nwpu-aslp.org (Sining Sun), lxie@nwpu.edu.cn (Lei Xie).}

\thanks{Linhao Dong is with Bytedance Speech, Beijing Bytedance Technology Co Ltd, Beijing 100098, China. Email: donglinhao@bytedance.com.}
}

%

%

\markboth{Journal of \LaTeX\ Class Files,~Vol.~14, No.~8, August~2015}%
{Wang \MakeLowercase{\textit{et al.}}: Bare Demo of IEEEtran.cls for IEEE Communications Society Journals}
%



\maketitle

\begin{abstract}
Data2vec is a self-supervised learning (SSL) approach that employs a teacher-student architecture for contextual representation learning via masked prediction, demonstrating remarkable performance in monolingual ASR.
Previous studies have revealed that data2vec's shallow layers capture speaker and language information, middle layers encode phoneme and word features, while deep layers are responsible for reconstruction. Language and phoneme features are crucial for multilingual ASR. However, data2vec's masked representation generation relies on multi-layer averaging, inevitably coupling these features.
To address this limitation, we propose a decoupling quantization based data2vec (DQ-Data2vec) for multilingual ASR, which includes a data2vec backbone and two improved online K-means quantizers.
Our core idea is using the K-means quantizer with specified cluster numbers to decouple language and phoneme information for masked prediction.
Specifically, in the language quantization, considering that the number of languages is significantly different from other irrelevant features (e.g., speakers), we assign the cluster number to match the number of languages, explicitly decoupling shallow layers' language-related information from irrelevant features. 
This strategy is also applied to decoupling middle layers' phoneme and word features.
In a self-supervised scenario, experiments on the CommonVoice dataset demonstrate that DQ-Data2vec achieves a relative reduction of ${9.51\%}$ in phoneme error rate (PER) and ${11.58\%}$ in word error rate (WER) compared to data2vec and UniData2vec. 
Moreover, in a weakly-supervised scenario incorporating language labels and high-resource language text labels, the relative reduction is ${18.09\%}$ and ${1.55\%}$, respectively.
\end{abstract}


\section{Introduction}
\label{sec:Introduction}
\IEEEPARstart{G}{eneral} end-to-end automatic speech recognition (E2E ASR) systems require many thousands of hours of human-annotated speech recordings as training data~\cite{chen2021gigaspeech, zhang2022wenetspeech}. 
However, in the multilingual scenario, apart from a few high-resource languages such as English and Chinese, most languages are low-resource~\cite{katzner2002languages}, constrained by the cost of recording or the number of speakers, making it difficult to obtain sufficient labeled data. 
In recent years, self-supervised learning (SSL)-based multilingual ASR approaches\cite{babu2021xlsr, chiu2022self, wang2021unispeech, xue2023unidata2vec} have achieved remarkable performance on phoneme error rate (PER) or word error rate (WER), by a pre-training process which without relying on target-language text labels, attracting significant attention. 
Specifically, Based on the data employed in pre-training, these multilingual SSL approaches can be classified into two types: (a) typical self-supervised learning without relying on any labels, such as XLS-R~\cite{babu2021xlsr} and BEST-RQ~\cite{chiu2022self}; (b) weakly supervised learning that additionally incorporates non-target language text labels, such as UniSpeech~\cite{wang2021unispeech} and UniData2vec~\cite{xue2023unidata2vec}.

Speech SSL approaches employ a two-stage process: pre-training on extensive unlabeled data, followed by fine-tuning on limited labeled datasets. In the multilingual scenario, this strategy enables the extraction of shared pronunciation features across languages by leveraging unlabeled data from non-target languages, ultimately enhancing ASR performance in the target language.
Data2vec~\cite{baevski2022data2vec} is a representative of these speech SSL approaches.
Data2vec is based on a teacher-student framework, where the teacher is responsible for generating target speech representations for masked frame tokens, and the student is then encouraged to reconstruct these target representations from the unmasked context. Notably, the target speech representation is derived from the average of the top K layers' results of the teacher branch.
The teacher branch mirrors the student's architecture and is updated from the student's parameters via the EMA~\cite{baevski2022data2vec,he2020momentum} method. 
This ensures that the teacher possesses the same context-understanding ability as the student, resulting in continuous and contextual target speech representations.
In contrast to discrete speech representations extracted from shallow features (e.g., wav2vec 2.0~\cite{baevski2020wav2vec} and vq-wav2vec~\cite{baevski2020vq}), these continuous and contextual representations are better suited for ASR tasks, leading to superior ASR performance~\cite{yang2021superb}.

Despite the proven efficacy of data2vec in multilingual ASR, its model architecture is designed for monolingual scenarios and underutilizes some helpful information for multilingual ASR tasks, such as cross-lingual shared phonemes and language identification.
Research has demonstrated that full utilization of such information causes further enhancement of multilingual ASR~\cite{pratap2020massively,liu2021unified,chen2023improving}. However, in data2vec, an unsupervised generation of language- and phoneme-related speech representations for masked frames faces challenges.
Previous research~\cite{pasad2021layer, shah2021all, xue2023sshr} showed that data2vec's shallow layers capture speaker- and language-related information, middle layers encompass phoneme- and word-related information, and deep layers are responsible for reconstruction.
However, the averaging operation of the teacher's top K layers couples phoneme- and language-related information.
Moreover, even if shallow and middle outputs are averaged separately, the coupling persists due to other unrelated details (e.g., speakers and words) still residing within the same layer, rendering the unsupervised decoupling of phoneme- and language-related information challenging.

In this paper, we propose \textbf{D}ecoupling \textbf{Q}uantization based Data2vec (DQ-Data2vec) for multilingual ASR tasks, which includes a data2vec backbone and two improved online K-means quantizers.
Our core idea is using the K-means quantizer with specified cluster numbers to decouple language and phoneme information for masked prediction.
Specifically, in a self-supervised scenario, firstly our DQ-Data2vec leverages data2vec as a backbone to extract shallow and middle features from the teacher branch. This ensures the incorporation of both language and phoneme characteristics in the extracted features.
Then, we further improve the typical online K-means vector quantizer~\cite{baevski2020vq, van2017neural} to decouple language and phoneme characteristics from unrelated details. 
In the language quantization, considering that the number of languages is significantly different from other irrelevant features, the cluster number is then specified to correspond to languages, thus explicitly decoupling language-related from language-irrelevant information (e.g., speakers). This strategy is also adopted for decoupling phonemes and words.
Finally, contrastive losses are employed to encourage the student branch to reconstruct the language- and phoneme-related speech representations.
Furthermore, in a weakly-supervised scenario, language labels and non-target language text labels are incorporated through supervised losses, simultaneously constraining the quantizer's outputs and the student branch's results, resulting in clearer language and phoneme representations to enhance SSL performance further.
Our experiments on the multilingual CommonVoice dataset~\cite{ardila2020common} demonstrate the effectiveness of the proposed model.
In the self-supervised and weakly-supervised scenarios, our DQ-Data2vec achieves a relative reduction of ${9.51\%}$/${18.09\%}$ in PER and ${11.58\%}$/${1.55\%}$ in WER compared to data2vec~\cite{baevski2022data2vec} and UniData2vec~\cite{xue2023unidata2vec} baselines, respectively.

The main contributions of this paper can be summarized as follows: 
\begin{itemize}
    \item We propose an SSL framework for multilingual ASR tasks, which explicitly extract language- and phoneme-related speech representations of masked frames, leading to superior performance in both self-supervised and weakly-supervised scenarios.
    \item We improve typical K-means quantizers to cluster specific target representations and decouple irrelevant information with expert knowledge, including specifying layer positions for quantization and aligning cluster numbers with the type counts of the quantizing targets.
    \item Our DQ-Data2vec approach demonstrates its flexibility. The decoupling of frame-level or utterance-level objects, guided by layer positions and targets' type counts, holds promise for application to other objects, such as speakers or emotions, thus guiding SSL to adapt to downstream tasks related to them.
\end{itemize}

\section{Related Works}
\label{sec:Related Works}
In this section, we present a summary that is relevant to this research, which includes unsupervised and weakly-supervised representation learning, and the decoupling approaches of specific targets in SSL.

\subsection{Self-supervised Speech Representation Learning}
The dominant speech SSL models~\cite{baevski2020wav2vec, hsu2021hubert, baevski2022data2vec, babu2021xlsr, chiu2022self, chen2022wavlm, chung2021w2v} follow a BERT-style framework, masking partial input tokens and then reconstruct them based on the remaining contextual inputs.
These methodologies have demonstrated superiority compared to earlier techniques~\cite{baevski2020vq, schneider2019wav2vec}, which involve predicting future tokens from past tokens. However, originally designed for NLP tasks, BERT SSL~\cite{devlin2018bert} always faces the challenge of token generation for unclear-boundary speech signals. Wav2vec 2.0~\cite{baevski2020wav2vec} generates discrete speech tokens using a quantizer with Gumbel softmax~\cite{jang2016categorical, maddison2014sampling}. W2v-BERT~\cite{chung2021w2v} builds upon wav2vec 2.0 and incorporates a masked language modeling (MLM) block, resulting in further enhancement. HuBERT~\cite{hsu2021hubert} generates pseudo-labels for speech tokens using offline K-means clustering. Data2vec~\cite{baevski2022data2vec} introduces an additional teacher branch to generate continuous context-related representations as tokens. Furthermore, BEST-RQ~\cite{chiu2022self} employs random quantization to generate IDs for speech tokens, with the mapping fixed during training. Generally, the above-mentioned speech SSL models necessitate distinguishable tokens at different frames, naturally associating them with phonemes. Therefore, when these methods are applied to multilingual scenarios~\cite{babu2021xlsr, chiu2022self}, they overlook utterance-level language identification information, while it plays a crucial role in enhancing multilingual ASR performance~\cite{pratap2020massively, liu2021unified, chen2023improving}. In contrast to existing self-supervised speech representation learning approaches, our proposed DQ-Data2vec, with shallow decoupling, can concurrently learn phoneme- and language-related information without any labels, thereby demonstrating specific advantages in multilingual speech tasks.

\subsection{Weakly-supervised Speech Representation Learning}
Weakly-supervised speech representation learning approaches~\cite{wang2021unispeech, xue2023unidata2vec, bai2022just, zhang2022xlst} have been derived from SSL and trained with limited labeled data of non-target high-resource languages and unlabeled data of target low-resource languages. Due to the presence of similar pronunciation units across different languages, these approaches can acquire shared phonetic information from high-resource languages, thereby benefiting target low-resource languages. In the aforementioned approaches, UniSpeech~\cite{wang2021unispeech} is derived from wav2vec 2.0~\cite{baevski2020wav2vec}, which mixes the quantization vectors and the outputs of the Transformer~\cite{vaswani2017attention} backbone, and then utilizes connectionist temporal classification (CTC) loss~\cite{graves2006ctc} to align them with the phoneme labels. UniData2vec~\cite{xue2023unidata2vec} is derived from data2vec~\cite{baevski2022data2vec} and has a similar design thinking to the UniSpeech, but all of their outputs come from the Transformer backbone without the mixture. These two approaches necessitate two-stage pre-training on high-resource and low-resource languages, respectively. JUST~\cite{bai2022just} inherits from w2v-BERT and adds a decoder with an RNN-T loss~\cite{graves2012rnnt}, which only involves one-stage pre-training but requires that all data have labels. XLST~\cite{zhang2022xlst} achieves competitive performance by replacing the teacher branch in data2vec~\cite{baevski2022data2vec} with a well-trained model. 
In contrast to these approaches, our proposed DQ-Data2vec with deep decoupling achieves superior performance with only a single stage of pre-training.

\subsection{Decoupling Specific Target for SSL}
The prevailing methods primarily steer SSL models to acquire language or speaker information by incorporating labeled data during the fine-tuning stage~\cite{yang2021superb, fan2020exploring, cho2022non, chen2023improving}. Limited research has endeavored to learn specific non-ASR information during the pre-training stage~\cite{chen2022wavlm, chen2022unispeech_sat}. In WavLM~\cite{chen2022wavlm}, Chen~\textit{et al.} combine masked speech prediction and denoising in pre-training, thereby the model learns speaker-related tasks through speech denoising, such as speaker verification, speech separation, and speaker diarization. In UniSpeech-SAT~\cite{chen2022unispeech_sat}, a quantizer with Gumbel softmax~\cite{baevski2020wav2vec} is integrated into the intermediate layer of HuBERT~\cite{hsu2021hubert} to extract speaker information. Specifically, utterance-level contrastive loss and utterance mixing augmentation render the information learned by the quantizer speaker-related. 
UniSpeech-SAT achieves excellent performance on speaker-related tasks with a light ASR performance reduction on the SUPERB~\cite{yang2021superb} benchmark.
In contrast to the methods above, the DQ-Data2vec model we propose facilitates the explicit extraction of specified frame-level and utterance-level information concurrently and leverages the complementarity of the two types of information to enhance performance on the multilingual ASR task.

\section{Method}
\label{sec:Method}

\begin{figure*}[htp]
\centering
\vspace{0em}
\includegraphics[width=45em]{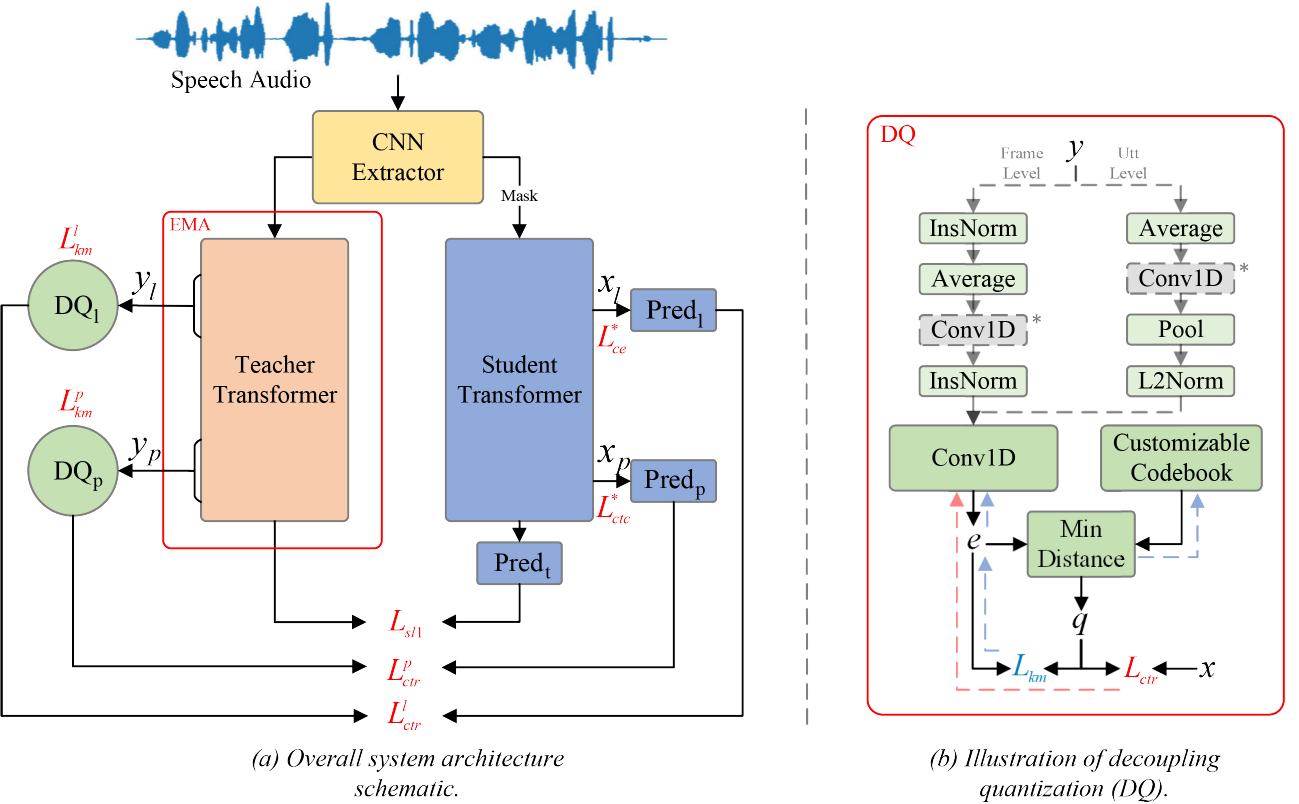}
\vspace{0em}
\caption{The framework overview of our proposed DQ-Data2vec. The symbol ${\ast}$ denotes that the object is only employed in the deep decoupling scenario. In the right figure, the gray dashed lines represent different designs for frame-level and utterance-level quantization, while the red and blue dashed lines represent the gradient back-propagation paths of different loss functions.}
\label{fig:system_framework}
\vspace{-1em}
\end{figure*}

Our proposed DQ-Data2vec is a BERT-style~\cite{devlin2018bert} SSL model, involving a teacher-student backbone and two online K-means vector quantizers~\cite{baevski2020vq, van2017neural} which are used to extract language and phoneme-related speech representations. It aims to enhance multilingual ASR performance by producing target speech representations with more transparent physical interpretations. The overall architecture is shown in Fig.~\ref{fig:system_framework}. These components of DQ-Data2vec will be further explained in the subsequent subsections.

\subsection{Teacher-Student Backbone}
\label{sec:teacher_student_backbone}
The data2vec~\cite{baevski2022data2vec} is an outperforming speech SSL, which is based on a teacher-student framework. Its teacher branch can express diverse and unmasked information across various layers. 
This multifaceted information will subsequently serve as the foundation for decoupling quantization with varying content.
Moreover, SSL is a precarious balance that is prone to collapse during pre-training~\cite{cao2023trinet}, this vulnerability will be particularly pronounced after the introduction of multiple quantizers.
Particularly in the initial stages of training when the physical significance of the intermediate layer outputs in the teacher branch is not yet apparent, the quantizer may yield low-quality speech representations. The data2vec backbone can stabilize the training process.
Therefore, we develop our multilingual SSL based on data2vec.
Specifically, this framework incorporates a convolutional feature encoder, a teacher Transformer encoder, and a student Transformer encoder. 
The parameters of the teacher encoder are updated from the student encoder using the EMA method~\cite{baevski2022data2vec} with stops all gradient back-propagation.
Initially, the convolutional feature encoder maps raw speech audio to latent representations. Then, the masked latent representations are put into the student branch, while the raw latent representations without masks are fed into the teacher branch. 
Following this, the results of the top ${K}$ layers in the teacher branch undergo instance normalization~\cite{ulyanov2016instance} and are then averaged to form a target speech representation, which is denoted as ${\mathbf{y}_{t}}$. 
Finally, the output of the student branch's last layer, ${\mathbf{x}_{t}}$, is transformed into ${\mathbf{x}^{'}_{t}}$ by a linear predictor ${\text{Pred}_{t}}$, and the masked frames in ${\mathbf{x}^{'}_{t}}$ are optimized to reconstruct the corresponding frames in ${\mathbf{y}_{t}}$ by using the Smooth L1
(SL1)~\cite{baevski2022data2vec} loss.
In the specific setup, the convolutional feature encoder comprises 7 temporal convolution layers, each with 512 channels. The strides are (5, 2, 2, 2, 2, 2, 2), and the kernel widths are (10, 3, 3, 3, 3, 2, 2). The two Transformer context encoders are comprised of 12 layers, with a model dimension of 768, an inner dimension of 3072, and attention heads of 8. In addition, a convolutional layer with a kernel size of 128 and 16 groups is used to replace absolute positional embedding. The top 8 layers are employed to produce ${\mathbf{y}_{t}}$. The predictor ${\text{Pred}_{t}}$ is a single linear layer without normalization or activation. The computation of the SL1 loss ${ L_{sl1}}$ can be denoted as:
\begin{equation}
\label{eq:pred_t}
\left\{\begin{array}{l}
  \mathbf{y}_{t}=\frac{1}{K}\sum_{i=L-K+1}^{L}\text{InsNorm}(\mathbf{y}_{i}) \\ 
  \mathbf{x}^{'}_{t}=\text{Pred}_{t}(\mathbf{x}_t)
\end{array}\right.,
\end{equation}
\begin{equation}
\label{eq:loss_sl1}
    L_{sl1}=\begin{cases} \frac{1}{2}(\mathbf{y}_{t}-\mathbf{x}^{'}_{t})^2/\beta & \text{ if } |\mathbf{y}_{t}-\mathbf{x}^{'}_{t}| \le \beta  \\ |\mathbf{y}_{t}-\mathbf{x}^{'}_{t}|-\frac{1}{2}\beta & \text{ otherwise}\end{cases},
\end{equation}
where ${L}$ represents the layer number of the Transformer encoders, ${K}$ denotes the top layer count of the teacher branch, specifically with ${L=12}$ and ${K=8}$ in this study, while $\beta$ is tunable and controls the transition from square loss to L1 loss.

\subsection{Shallow Decoupling}
\label{sec:shallow_decoupling}
In previous research~\cite{pasad2021layer,shah2021all}, it has been observed that the results from different layers of the SSL model's backbone encompass distinctive information. 
Specifically, shallow layers capture speaker- and language-related information, middle layers encompass phoneme- and word-related information, and deep layers are responsible for restructuring SSL's masked speech representations.
The speech representations extracted from specific layers are expected to guide SSL models toward adapting to diverse tasks. However, one layer's results may encompass multiple types of information. For example, language and speaker information may coexist within the same shallow layer. 
Therefore, directly utilizing the specific layer's results as the target speech representation is insufficient for enabling the SSL model to capture the intended content.
In this study, we introduce two online K-means quantizers~\cite{baevski2020vq,van2017neural} to extract speech representations from the teacher branch, which allows decoupling specific targets by regulating the number of cluster centers and the utilization of pooling.
Specifically, when seeking to extract utterance-level language-related speech representations, the introduction of temporal pooling can eliminate frame-level information, and further adjusting the number of cluster centers to match the number of languages can substantially decouple speaker information. 
Notably, the goal of setting the cluster number to language count is not to establish a one-to-one mapping between the quantitative representation and languages, which is impractical due to the absence of labels' constraints. Our motivation lies in leveraging the significant difference between the number of speakers and languages to make the clustering results as close as possible to the languages.
Moreover, in the case of requiring a frame-level phoneme-related target speech representation, the intermediate layer results can be quantified with the number of cluster centers set to the phoneme count, and temporal pooling can be omitted.
In this study, we refer to this scenario as shallow decoupling, which does not rely on any labeled data and only specifies the layer position, the number of cluster centers, and whether pooling is used to specify the content of the speech representation. Below, we detail the designs in the shallow decoupling scenario.

\textbf{Layer Results.} 
As shown in Fig.~\ref{fig:system_framework}, we select the shallow layers' results ${\mathbf{y}_{l}}$ from the teacher branch for language quantization, and the intermediate layers ${\mathbf{y}_{p}}$ for phoneme quantization. Based on the findings in~\cite{pasad2021layer}, we designate ${\mathbf{y}_{l}=\left\{4,5,6\right\}}$ layers, and ${\mathbf{y}_{p}=\left\{7,8,9\right\}}$ layers. 
Considering these two quantizations correspond to utterance-level and frame-level, respectively, ${\mathbf{y}_{l}}$ and ${\mathbf{y}_{p}}$ require distinct pre-processing to manifest the level-related characteristics. Specifically, in our study, the level-related pre-processing exclusively involves normalization, pooling, and averaging.
It is noteworthy that no learnable parameters are involved in the pre-processing, as we have experimentally determined that the incorporation of excessive learnable parameters following the teacher-branch layers' outputs before quantization, may result in a collapse during self-supervised training.
The specific pre-processing settings are as follows:

\begin{itemize}
    \item Utterance-level pre-processing: 
    Language-specific information is only apparent after a temporal pooling. This characteristic typically emerges across various feature dimensions within an utterance and is independent of other utterances within the same batch. Consequently, L2 normalization~\cite{goodfellow2016deep} is appropriate for this scenario as it operates on the feature dimension. Illustrated in Fig.~\ref{fig:system_framework} (b), the utterance-level pre-processing can be denoted as:
    \begin{equation}
    \label{eq:yl_}
        \mathbf{y}^{'}_{l}=\text{L2Norm}(\text{Pool}(\frac{1}{N_{l}}\sum\mathbf{y}_{i})), \mathbf{y}_{i}\in\mathbf{y}_{l},
    \end{equation}
    where ${N_{l}}$ is the layer count of ${\mathbf{y}_{l}}$.
    
    \item Frame-level pre-processing: 
    Phoneme-related information is typically context-dependent. Normalizing the same feature dimension at different time steps within an utterance can effectively capture the distinctions between various frames. Hence, instance normalization~\cite{ulyanov2016instance} is appropriate for this scenario. As shown in Fig.~\ref{fig:system_framework} (b), the frame-level pre-processing can be denoted as:
    \begin{equation}
    \label{eq:yp_}
       \mathbf{y}^{'}_{p}=\text{InsNorm}(\frac{1}{N_{p}}\sum\text{InsNorm}(\mathbf{y}_{i})),\mathbf{y}_{i}\in\mathbf{y}_{p},
    \end{equation}
    where ${N_{p}}$ is the layer count of ${\mathbf{y}_{p}}$.
\end{itemize}
In addition, it is worth noting that the selection of normalization for various levels here is imperative for the stability and efficacy of our DQ-Data2vec framework. We have explored the effect of normalization mismatch by using instance normalization for utterance-level pre-processing and L2 normalization for frame-level pre-processing. The former significantly reduced the language-related information in the quantized results, while the latter caused a loss explosion during training.

\textbf{Online K-means Quantizer.}
In~\cite{van2017neural}, van den Oord~\textit{et al.} introduced the online K-means quantizer for SSL models. This method maintains a finite set of codewords, known as a codebook, which is randomly initialized and can be updated through back-propagation. 
Within the codebook, the codeword ${\mathbf{c}_{i}}$ that has the smallest Euclidean distance from the input vector ${\mathbf{e}}$ is chosen as the quantization result ${\mathbf{q}}$. This selection is represented by Equation~\ref{eq:codeword_selection}:
\begin{equation}
\label{eq:codeword_selection}
\left\{\begin{array}{l}
  i=\arg\min_{j}\left \|\mathbf{e}-\mathbf{c}_j\right \|^{2}, \mathbf{c}_{j}\in\mathbb{R}^{N\times D} \\ 
  \mathbf{q}=\mathbf{c}_i
\end{array}\right.,
\end{equation}
where ${N}$ represents the number of codewords and ${D}$ denotes the feature dimension of the input vector. In this study, ${N}$ is set to the number of languages and phonemes, respectively.
During the back-propagation process, the mean squared error (MSE) loss is utilized to successively guide the codebooks and the input vectors to move toward each other.
The loss function for the online K-means quantization is represented by Equation~\ref{eq:loss_km}:
\begin{equation}
    \label{eq:loss_km}
    L_{km}=\left \|\text{sg}(\mathbf{e})-\mathbf{q}\right \|^{2}+\gamma\left \|\mathbf{e}-\text{sg}(\mathbf{q})\right \|^{2},  
\end{equation}
where ${\text{sg}}$ denotes the stop gradient operation and ${\gamma}$ represents a tunable hyperparameter (${\gamma\ne1.0}$, with a default value of ${\gamma=0.25}$). 
The use of ${\text{sg}}$ and ${\gamma}$ enables the asynchronous updating of the codebook and input vector to prevent collapse. 
In our study, all layers' outputs in the teacher branch are subjected to gradient stopping when the EMA method is used. 
Hence, if ${\mathbf{y}^{'}_{l}}$ (Equation~\ref{eq:yl_}) and ${\mathbf{y}^{'}_{p}}$ (Equation~\ref{eq:yp_}) are directly utilized as the input vectors ${\mathbf{e}}$ for the quantizers, the second term in Equation~\ref{eq:loss_km} cannot update any parameters.
To address this, as illustrated in the green Conv1D block in Fig.~\ref{fig:system_framework} (b), we introduce a temporal convolution layer for ${\mathbf{y}^{'}_{l}}$ and ${\mathbf{y}^{'}_{p}}$, i.e., ${\mathbf{e}=\text{Conv1D}(\mathbf{y}^{'})}$, which gives significance to the second term in Equation~\ref{eq:loss_km}. 
It is noteworthy that the stride, kernel width, and groups in this convolution are set to 1, 1, and 2. The first two parameters are intended to reduce the contextual information learning ability of the convolutional layer to avoid training collapse.
The group parameters are set to 2 to simulate product quantization~\cite{jegou2010product}. 
This approach divides the feature dimension into two segments and sequentially quantizes them, which is a strategy demonstrated to significantly boost quantizer performance~\cite{baevski2020wav2vec}.

\textbf{Quantization Learning Objective.} 
In our proposed DQ-Data2vec, there are three target speech representations ${\mathbf{y}_{t}}$ (Equation~\ref{eq:pred_t}), ${\mathbf{q}_{l}}$, and ${\mathbf{q}_{p}}$ (Equation~\ref{eq:codeword_selection}). The student branch needs to simultaneously learn to reconstruct these representations. 
Where ${\mathbf{y}_{t}}$ is reconstructed from the student encoder's final layer ${\mathbf{x}_{t}}$. 
For ${\mathbf{q}_{l}}$ and ${\mathbf{q}_{p}}$, we reconstruct them by the student layers ${\mathbf{x}_{l}}$ and ${\mathbf{x}_{p}}$, which are positioned at the deepest layers in ${\mathbf{y}_{l}}$ and ${\mathbf{y}_{p}}$, specifically ${\mathbf{x}_{l}=6}$ and ${\mathbf{x}_{p}=9}$.
Additionally, similar to ${\mathbf{x}_{t}}$, we establish two predictors, ${\text{Pred}_{l}}$ and ${\text{Pred}_{p}}$, for mapping ${\mathbf{x}_{l}}$ and ${\mathbf{x}_{p}}$. The mapping vectors ${\mathbf{x}^{'}_{l}}$ and ${\mathbf{x}^{'}_{p}}$ can be represented as follows:
\begin{equation}
\label{eq:pred_lp}
\left\{\begin{array}{l}
  \mathbf{x}^{'}_{l}=\text{Pool}(\text{Pred}_{l}(\mathbf{x}_l)) \\
  \mathbf{x}^{'}_{p}=\text{Pred}_{p}(\mathbf{x}_p)
\end{array}\right.,
\end{equation}
where ${\text{Pred}_{l}}$ and ${\text{Pred}_{p}}$ are composed of 2 Transformer layers and 1 linear layer. Upon obtaining ${\mathbf{x}^{'}_{l}}$ and ${\mathbf{x}^{'}_{p}}$, we employ contrastive loss~\cite{baevski2020wav2vec} for the reconstructing. The quantized vectors produced by the online K-means serve as the clustering centers of the teacher branch's embeddings, encompassing rich language and phoneme-related details. Utilizing the contrastive loss can steer the distribution of feature dimensions in ${\mathbf{x}^{'}_{l}}$ and ${\mathbf{x}^{'}_{p}}$ to align with ${\mathbf{q}_{l}}$ and ${\mathbf{q}_{p}}$, thereby incorporating these details into the student branch. Furthermore, according to the research in~\cite{chen2021exploring}, contrastive learning introduces additional quantized vectors as negative examples, which is advantageous for mitigating training collapse.
The contrastive loss is defined as follows:
\begin{equation}
    \label{eq:loss_ctr}
    L_{ctr}=-\log\frac{\exp{(\text{Sim}(\mathbf{x}^{'},\mathbf{q})/\kappa)}}{ \sum_{\mathbf{\widehat{q}}\sim{\mathbf{Q}}}\exp{(\text{Sim}(\mathbf{x}^{'},\mathbf{\widehat{q}})/\kappa)}} ,
\end{equation}
where ${\text{Sim}}$ represents the cosine similarity, ${\kappa}$ denotes a non-negative temperature (default ${\kappa=0.1}$), and ${\mathbf{Q}}$ stands for the negative examples. 
We choose other masked frames of the intra-utterance as negative examples for the phoneme quantizer. Regarding the language quantizer, the negative examples are chosen from inter-utterance. It is important to note that, as depicted in Fig.~\ref{fig:system_framework} (b), we stop the gradient from ${L_{ctr}}$ to the codebook and solely update the temporal convolution layer to mitigate disturbance during clustering. Ultimately, the loss of quantization learning  can be expressed as:
\begin{equation}
\label{eq:loss_qt}
  L_{qt}=L_{ctr}+L_{km} , \\
\end{equation}

In the shallow decoupling scenario, the total loss can be denoted as follows:
\begin{equation}
\label{eq:loss_unlabel}
  L_{sc}=(1-\gamma_{1}-\gamma_{2})L_{sl1}+\gamma_{1}L^{l}_{qt}+\gamma_{2}L^{p}_{qt} ,
\end{equation}
where ${\gamma_{1}}$ and ${\gamma_{2}}$ are tunable hyperparameters, which are set to 0.1 and 0.2 in this study.

\subsection{Deep Decoupling}
The shallow decoupling of language and phoneme quantization established using unlabeled data is a weak constraint, with limited accuracy. The introduction of labeled data is expected to augment the quality of quantized speech representations. In comparison to text labels for low-resource languages, language labels and text labels of non-target high-resource languages are more easily obtainable. Leveraging these labels can further enhance the performance of our approach without significantly increasing the cost. Drawing from UniSpeech~\cite{wang2021unispeech} and UniData2vec~\cite{xue2023unidata2vec}, we designate the scenario as deep decoupling, which involves utilizing all language labels and text labels of non-target high-resource languages.

In the deep decoupling scenario, 
we mix the outputs of the student layers ${\mathbf{x}}$ and the quantized vectors ${\mathbf{q}}$ in a 1:1 ratio to create unified representations, which denoted as ${\mathbf{u}}$. 
Specifically, for the phoneme unified representation ${\mathbf{u}_{p}}$, we randomly mask ${50\%}$ of the frames in $\mathbf{x}_p$ within each utterance, replacing them with corresponding frames from $\mathbf{q}_p$ to create a mixture.
The language unified representation $\mathbf{u}_l$ is constructed similarly, but instead of frame-level replacement within an utterance, it uses utterance-level replacement within a batch.
The unified representations $\mathbf{u}_l$ and $\mathbf{u}_p$ can be denoted as:
\begin{equation}
\label{eq:unified}
\left\{\begin{array}{l}
  \mathbf{u}_{l}=\text{Mix}(\text{Pool}(\mathbf{x}_l),\mathbf{q}_l)=\text{Pool}(\mathbf{x}_l)*\mathbf{M}_l+\mathbf{q}_l*(1-\mathbf{M}_l) \\
  \mathbf{u}_{p}=\text{Mix}(\mathbf{x}_p,\mathbf{q}_p)=\mathbf{x}_p*\mathbf{M}_p+\mathbf{q}_p*(1-\mathbf{M}_p)
\end{array}\right.,
\end{equation}
where ${\mathbf{M}_l}$ and ${\mathbf{M}_p}$ are binary mask matrices, ${\mathbf{M}_l\in\mathbb{R}^{1\times D}}$ and ${\mathbf{M}_p\in\mathbb{R}^{T\times D}}$
. In addition, ${\mathbf{x}_{l}}$ and ${\mathbf{x}_{p}}$ contain the masked and unmasked frames.

Following this, the cross entropy (CE) loss and CTC loss are employed to instruct the model in learning language and phoneme information, respectively. The computation of the supervised loss functions is represented as follows:
\begin{equation}
\label{eq:loss_ce}
  L_{ce}=\text{CrossEntropy}(\mathbf{u}_{l}, \mathbf{Y}_{l}) ,
\end{equation}
\begin{equation}
\label{eq:loss_ctc}
    L_{ctc}=\begin{cases} \text{CTC}(\mathbf{u}_{p}, \mathbf{Y}_{p}) & \text{ if } \mathbf{Y}_{p} \in\mathbb{H}  \\ 
    0 & \text{ otherwise } \end{cases},
\end{equation}
where ${\mathbf{Y}_{l}}$ and ${\mathbf{Y}_{p}}$ refer to language and phoneme text labels, respectively. ${\mathbb{H}}$ is the non-target high-resource languages.

The mixture enhances the quality of the quantized representation both directly and indirectly. The former involves optimizing quantizers through gradient back-propagation of ${\mathbf{q}_{l}}$ and ${\mathbf{q}_{p}}$, while the latter entails optimizing the student branch using the gradient from ${\mathbf{x}_{l}}$ and ${\mathbf{x}_{p}}$, and then updating the teacher using EMA, facilitating easier clustering of ${\mathbf{y}_{l}}$ and ${\mathbf{y}_{p}}$. However, if the quantizer maintains the same design as in shallow decoupling, the gradient of ${\mathbf{q}_{l}}$ and ${\mathbf{q}_{p}}$ will be stopped by EMA and only update the green temporal convolution block in Fig.~\ref{fig:system_framework} (b). As mentioned in Section~\ref{sec:Method}.B, this convolution layer's kernel size is set to 1 to split the feature dimension into two parts and simulate product quantization. This configuration makes the convolution less effective in capturing the timing-related language and phoneme information, thus not fully utilizing the gradient of the supervised loss. Therefore, we additionally introduce a convolution block (the grey block in Fig.~\ref{fig:system_framework} (b)) with a larger kernel size in the deep decoupling scenario to further emphasize the language and phoneme information in ${\mathbf{y}_{l}}$ and ${\mathbf{y}_{p}}$, thereby enhancing the clustering effect. Specifically, this additional convolution block consists of 2 layers, and the kernel width is set to 3. It is important to note that this block can only be used when the supervised loss is employed, and introducing it in shallow decoupling would result in training collapse.

Finally, in the deep decoupling scenario, the total loss can be denoted as follows:
\begin{equation}
\label{eq:loss_unlabel}
  L_{dc}=L_{sc}+\gamma_{3}(L_{ce}+L_{ctc}) ,
\end{equation}
where ${\gamma_{3}}$ is set to 0.1 in this study.

\section{Experiments Setup}
\label{sec:Experiments Setup}
\subsection{Dataset}
We conduct comprehensive experiments on the publicly-available multilingual dataset CommonVoice 6.0\footnote{Accessible at \url{https://commonvoice.mozilla.org}}~\cite{ardila2020common} to evaluate the ASR performance of our proposed DQ-Data2vec. To ensure comparability with prior studies~\cite{babu2021xlsr,wang2021unispeech,xue2023unidata2vec}, we conduct experiments using 9 languages: English (en), Spanish (es), French (fr), Italian (it), Kyrgyz (ky), Dutch (nl), Russian (ru), Tatar (tt), and Swedish (sv). Among these languages, en is a non-target high-resource language, while the others are target low-resource languages, and the experimental results are exclusively reported for the latter. The dataset details are presented in Table~\ref{tab:datasets}. 
We report both PER and WER results, with PER facilitating comparisons to prior studies~\cite{babu2021xlsr,wang2021unispeech,xue2023unidata2vec} and WER serving as a complementary metric.
During the fine-tuning for PER results, we utilize IPA phonemes as the modeling units, which are generated by the phonemizer\footnote{Accessible at \url{https://github.com/bootphon/phonemizer}} tool. 
Notably, all languages share a common dictionary comprising 169 phonemes and 5 special symbols.
In the fine-tuning for WER results, character units plus a space symbol are used, 
with the shared dictionary encompassing 298 units.
\begin{table}[htp]
    \centering
    \footnotesize
    \caption{Details of the CommonVoice datasets.}
    \label{tab:datasets}
    \begin{tabular}{@{}cccccccccc@{}}
    \toprule
    \multirow{2}{*}{\textbf{\begin{tabular}[c]{@{}c@{}}Sampling Rate\\ (kHz)\end{tabular}}} & \multicolumn{9}{c}{\textbf{Data Duration (hours)}} \\ \cmidrule(l){2-10} 
    & \textbf{en} & \textbf{es} & \textbf{fr} & \textbf{it} & \textbf{ky} & \textbf{tt} & \textbf{nl} & \textbf{ru} & \textbf{sv} \\ \midrule
    16 & 1350 & 168 & 353 & 90 & 17 & 17 & 29 & 55 & 3 \\ \bottomrule
    \end{tabular}
\end{table}

\subsection{Implementation Details}
Our experiments are conducted using the fairseq tool~\cite{ott2019fairseq}. In contrast to UniSpeech~\cite{wang2021unispeech} and UniData2vec~\cite{xue2023unidata2vec}, which necessitate separate two-stage pre-training on non-target high-resource languages and target low-resource languages, our DQ-Data2vec only requires single-stage pre-training on a mixture of these two data types, followed by fine-tuning on 1-hour labeled data.
During the pre-training of DQ-Data2vec, the settings for the time-step mask ratio, EMA, optimizer, and learning rate scheduler mirror those in the speech-processing data2vec~\cite{baevski2022data2vec}. 
Specifically, approximately ${49\%}$ of time steps are masked, and the optimizer follows the Adam~\cite{kingma2014adam} algorithm with a tri-stage scheduler that warms up, holds, and decays over the first ${3\%}$, the middle ${90\%}$, and the last ${7\%}$ of updates, respectively.
Moreover, recognizing that multilingual tasks pose greater difficulty than mono-lingual tasks in the context of data2vec, a large learning rate could result in gradient explosion. Therefore, we set the learning rate to ${3\times10^{-4}}$ across all our experiments. These models were trained for 400K updates on 16 A100-80g GPUs, with a maximum of 2500000 tokens in one batch.
During the fine-tuning phase, the models undergo 13K updates with a learning rate of ${5\times10^{-5}}$. The Transformer backbone is frozen during the first 10k updates, after which all parameters are updated collectively in the final 3k steps.

Moreover, the contrastive loss in quantization learning requires a varied array of samples within a batch, particularly for language diversity. However, the proportion of non-target English language data in Table~\ref{tab:datasets} is excessive, prompting the need for a data balance strategy during pre-training. To address this, we employ a specific resampling approach for speech utterances from a multinomial distribution ${p\sim(\frac{n_{l}}{N})^{0.5}}$, where ${n_{l}}$ denotes the number of pre-training hours for language l and ${N}$ represents the total hours~\cite{wang2021unispeech}. This method involves sampling an utterance multiple times in short-duration languages and randomly skipping a portion of utterances in a long-duration language.

\section{Experiments Results}
\label{sec:Experiments Results}
This section presents the experimental results of our proposed DQ-Data2vec. We first introduce the results of baselines and our approach in shallow decoupling and deep decoupling scenarios, respectively. Then, we conduct a detailed analysis of the effectiveness of the quantizer. Subsequently, we compare the effects of different designs in DQ-Data2vec.

\begin{table*}[htp]
\centering
\footnotesize
\caption{PER results on the CommonVoice dataset for our approach. The data balance strategy mentioned in Section~\ref{sec:Experiments Setup}.B is used by default. The table's DB, PQT, and LQT refer to the data balance, phoneme quantizer, and language quantizer. The gray background represents the experiment using the text labels of en language and all language labels. The models with references indicate their results come from previous research; otherwise, they are our own implementation.}
\label{tab:main_results}
\begin{threeparttable}
\begin{tabular}{@{}lccccccccc@{}}
\toprule
\multicolumn{1}{c}{\multirow{2}{*}{\textbf{Model}}} & \multicolumn{9}{c}{\textbf{PER (${\downarrow}$)}} \\ \cmidrule(lr){2-10}
\multicolumn{1}{c}{} & \textbf{es} & \textbf{fr} & \textbf{it} & \textbf{ky} & \textbf{tt} & \textbf{nl} & \textbf{ru} & \textbf{sv} & \textbf{Avg} \\ \hline\hline
\multicolumn{10}{l}{\textit{\textbf{Self-supervised Baseline}}} \\ \midrule
XLSR\tnote{\dag} w/o DB~\cite{wang2021unispeech} & 6.50 & 9.00 & 9.30 & 8.20 & 7.40 & 9.80 & 10.1 & 20.50 & 10.10 \\
wav2vec 2.0 & 8.16 & 10.76 & 11.21 & 8.45 & 8.20 & 15.28 & 14.93 & 20.97 & 12.24 \\
data2vec w/o DB & 5.86 & 7.63 & 8.04 & 8.63 & 7.14 & 12.42 & 11.51 & 18.27 & 9.94 \\
data2vec & 4.96 & 7.32 & 6.38 & 4.70 & 4.71 & 9.43 & 8.72 & 17.67 & 7.99 \\ \midrule
\multicolumn{10}{l}{\textit{\textbf{Weakly-supervised Baseline}}} \\ \midrule
\rowcolor[HTML]{EFEFEF} 
UniSpeech w/o DB~\cite{wang2021unispeech} & 5.70 & 7.90 & 8.10 & 6.80 & 6.00 & 9.30 & 8.60 & 17.70 & 8.76 \\
\rowcolor[HTML]{EFEFEF} 
UniSpeech & 8.79 & 11.01 & 11.17 & 9.00 & 8.45 & 14.45 & 14.15 & 19.81 &  12.10 \\
\rowcolor[HTML]{EFEFEF} 
UniData2vec w/o DB~\cite{xue2023unidata2vec} & 5.50 & 6.50 & 7.10 & 6.80 & 7.20 & 8.00 & 8.80 & 17.80 & 8.46 \\
\rowcolor[HTML]{EFEFEF} 
UniData2vec & 4.43 & 6.54 & 5.75 & 4.42 & 4.13 & \textbf{7.45} & 7.92 & 16.95 & 7.20 \\ \midrule
\multicolumn{10}{l}{\textit{\textbf{Shallow Decoupling (Self-supervised)}}} \\ \midrule
DQ-Data2vec w/o PQT & 4.68 & 6.61 & \textbf{5.76} & 4.59 & 4.48 & \textbf{8.12} & \textbf{7.88} & 16.32 & 7.31 \\
DQ-Data2vec w/o LQT & 4.96 & 7.29 & 6.26 & 4.57 & 4.38 & 8.77 & 8.30 & 16.49 & 7.63 \\
DQ-Data2vec & \textbf{4.54} & \textbf{6.50} & 5.99 & \textbf{4.33} & \textbf{4.20} & 8.42 & 8.02 & \textbf{15.80} & \textbf{7.23} \\ \midrule
\multicolumn{10}{l}{\textit{\textbf{Deep Decoupling (Weakly-supervised)}}} \\ \midrule
\rowcolor[HTML]{EFEFEF} 
DQ-Data2vec w/o PQT & 4.50 & \textbf{6.16} & 5.63 & 4.36 & 4.36 & 7.91 & 7.98 & 16.46 & 7.17 \\
\rowcolor[HTML]{EFEFEF} 
DQ-Data2vec w/o LQT & 4.36 & 6.40 & 5.61 & 4.34 & 4.05 & 7.73 & 7.66 & 15.92 & 7.01 \\
\rowcolor[HTML]{EFEFEF} 
DQ-Data2vec & \textbf{4.34} & 6.33 & \textbf{5.59} & \textbf{4.22} & \textbf{3.93} & 7.64 & \textbf{7.63} & \textbf{15.78} & \textbf{6.93} \\ \bottomrule
\end{tabular}
\begin{tablenotes}
    \centering
    \footnotesize
    \item{\dag}: This XLSR model is re-run in~\cite{wang2021unispeech} which trained on only the CommonVoice dataset, instead of the original version released by Conneau~\textit{et al.} in~\cite{brno2021xlsr}.
\end{tablenotes}
\end{threeparttable}
\end{table*}

\begin{table*}[htp]
\centering
\footnotesize
\caption{WER results on the CommonVoice dataset. SD and DD in the table refer to shallow decoupling and deep decoupling, respectively. The Avg3 and Avg8 in the table refer to the average WER on only \{es, fr, it\} languages and on all languages, respectively. The models with references indicate their results come from previous research; otherwise, they are our own implementation.}
\label{tab:wer_results}
\begin{threeparttable}
\begin{tabular}{@{}lcccccccccc@{}}
\toprule
\multicolumn{1}{c}{\multirow{2}{*}{\textbf{Model}}} & \multicolumn{10}{c}{\textbf{WER (${\downarrow}$)}} \\ \cmidrule(l){2-11} 
\multicolumn{1}{c}{} & \textbf{es} & \textbf{fr} & \textbf{it} & \textbf{ky} & \textbf{tt} & \textbf{nl} & \textbf{ru} & \textbf{sv} & \textbf{Avg3} & \multicolumn{1}{l}{\textbf{Avg8}} \\ \hline\hline
\multicolumn{11}{l}{\textit{\textbf{Self-supervised}}} \\ \midrule
data2vec & 31.54 & 45.39 & 27.15 & 24.55 & 23.90 & 40.46 & 41.27 & 46.94 & 34.69 & 35.15 \\
DQ-Data2vec (SD) & \textbf{26.25} & \textbf{39.30} & \textbf{24.53} & \textbf{20.71} & \textbf{19.61} & \textbf{33.63} & \textbf{37.29} & \textbf{47.29} & \textbf{30.03} & \textbf{31.08} \\ \midrule
\multicolumn{11}{l}{\textit{\textbf{Weakly-supervised}}} \\ \midrule
UniData2vec w/o DB~\cite{xue2023unidata2vec} & 29.5 & 45.3 & 29.6 & - & - & - & - & - & 34.8 & - \\
UniData2vec w/o DB + Transcoder\tnote{\dag}~\cite{xue2023unidata2vec} & 25.7 & \textbf{36.2} & 27.4 & - & - & - & - & - & 29.8 & - \\
UniData2vec & 24.07 & 39.57 & 23.55 & 21.02 & 24.84 & 31.93 & \textbf{37.08} & \textbf{45.38} & 29.06 & 30.93 \\
DQ-Data2vec (DD) & \textbf{23.63} & 39.08 & \textbf{23.12} & \textbf{20.33} & \textbf{19.74} & \textbf{31.60} & 37.34 & 49.35 & \textbf{28.61} & \textbf{30.52} \\ \bottomrule
\end{tabular}
\begin{tablenotes}
    \centering
    \footnotesize
    \item{\dag}: Transcoder is a phoneme-to-word translate model that proposed together with UniData2vec in~\cite{xue2023unidata2vec}, which trained on external text datasets without audio.
\end{tablenotes}
\end{threeparttable}
\end{table*}

\subsection{Shallow Decoupling}
In a self-supervised scenario, only unlabeled data are available. Therefore, we select wav2vec 2.0 and data2vec as baselines, which are the representative approaches of the online discrete quantization and teacher-student framework in SSL, respectively. 
Specifically, in Table~\ref{tab:main_results}, XLSR is a wav2vec 2.0-based model from previous research~\cite{wang2021unispeech}, which sequentially undergoes two-stage pre-training on non-target en language and other target multilingual data. 
In contrast, our proposed DQ-Data2vec involves only one-stage pre-training on a mixture of target and non-target language data with data balance. Thus, we supplement a wav2vec 2.0 baseline with data balance, as well as two data2vec baselines with and without data balance, and these three baselines undergo only one-stage pre-training and use the same learning rate as DQ-Data2vec. As depicted in Table~\ref{tab:main_results}, the PER of wav2vec 2.0 in the second row has significantly increased compared to XLSR w/o DB in the first row. This demonstrates that wav2vec 2.0 is insensitive to data balance, and the introduction of data balance does not compensate for the performance reduction caused by the abandonment of the second-stage pre-training on the target languages. In contrast, comparing the third and fourth rows, it can be observed that data2vec is sensitive to data balance, as even with only one-stage pre-training, data2vec can achieve a minimum average PER of 7.99 with a data balance strategy.

In Table~\ref{tab:main_results}, with regards to our DQ-Data2vec, when solely employing the language quantizer (DQ-Data2vec w/o PQT), the average PER decreased from the data2vec baseline's ${7.99\%}$ to ${7.31\%}$. This suggests that the language quantizer effectively extracts language information that assists the ASR task. When only the phoneme quantizer is used, DQ-Data2vec w/o LQT also achieved a PER reduction compared to the data2vec baseline, although the decrease was not as significant as with DQ-Data2vec w/o PQT. We think this is due to the greater difficulty of phoneme clustering compared to language clustering in the absence of labels under shallow decoupling. Additionally, for the BERT-style model, their speech representation tokens already encompass phoneme information, and the complementarity of language information is more obvious. 
Finally, when employing both the language and phoneme quantizers simultaneously, DQ-Data2vec achieves the best ASR performance with a PER of ${7.23\%}$, representing a relative ${9.51\%}$ reduction compared to the data2vec baseline. This demonstrates that language and phoneme quantization information are complementary, and their simultaneous use can lead to further ASR performance improvements compared to a single quantizer. These experiments illustrate the effectiveness of our proposed DQ-Data2vec in the shallow decoupling scenario without the use of any labeled data.

Furthermore, in Table~\ref{tab:wer_results}, we also show the character-level fine-tuning results with the WER metric. Comparing the results of the first and second rows in the table, it can be found that after the introduction of decoupling quantization, the WER has a significant decrease in all languages, and the average WER of all languages is relatively reduced by ${11.58\%}$.

\subsection{Deep Decoupling}
In a weakly-supervised scenario, we choose UniSpeech~\cite{wang2021unispeech} and UniData2vec~\cite{xue2023unidata2vec} as two comparable baselines. Their original applications entail two-stage pre-training without a data balance. 
In contrast, to compare with our DQ-Data2vec, we re-run the UniSpeech and UniData2vec with one-stage pre-training and data balance, setting the loss weights and learning rates to match those used for DQ-Data2vec. 
As depicted in Table~\ref{tab:main_results}, these baselines exhibit similar trends to those observed in the self-supervised baselines. 
Specifically, UniSpeech shows insensitivity to data balance, and even with its inclusion, UniSpeech fails to compensate for the performance decrease resulting from the abandonment of the second-stage pre-training. 
Conversely, UniData2vec demonstrates sensitivity to data balance, requiring only one-stage pre-training with data balance to surpass the original version of UniData2vec. Among these four results, the PER of ${8.46\%}$ achieved by UniData2vec w/o DB~\cite{xue2023unidata2vec} represents the current published state-of-the-art (SOTA) performance on the CommonVoice dataset. Our rerun of UniData2vec has surpassed this result, achieving a ${7.20\%}$ PER, and as a result, it has been selected as the comparison baseline in the deep decoupling scenario for this study.

In the results presented in Table~\ref{tab:main_results}, our DQ-Data2vec demonstrates a further improvement in multilingual ASR tasks compared to UniData2vec. Specifically, when utilizing language and phoneme quantizers, our DQ-Data2vec achieves a relative reduction of ${3.75\%}$ in PER compared to our rerun UniData2vec, and a relative reduction of up to ${18.09\%}$ compared to the original UniData2vec w/o DB~\cite{xue2023unidata2vec}. These results effectively illustrate the advantages of our DQ-Data2vec in deep decoupling scenarios. Furthermore, even without language information, the performance of DQ-Data2vec w/o LQT surpasses that of UniData2vec, indicating that the introduction of the decoupling quantization can more effectively utilize non-target language labels than directly adding CTC loss. Additionally, our DQ-Data2vec only necessitates one-stage pre-training, streamlining the training process. Moreover, in comparison to the shallow decoupling scenario, DQ-Data2vec demonstrates relative performance improvements of ${1.91\%}$, ${8.12\%}$, and ${4.15\%}$ with the help of labels. This shows that even in the absence of text labels in the target language, our DQ-Data2vec can leverage language labels and non-target language text labels to enhance ASR tasks in the target language. 
In addition, it is notable that, unlike in the shallow decoupling scenario, DQ-Data2vec w/o LQT exhibits greater strength than DQ-Data2vec w/o PQT in the deep decoupling scenario, indicating that the phoneme quantizer performs more effectively with the assistance of text labels and brings about a more pronounced auxiliary effect.

The results in Table~\ref{tab:wer_results} show that our DQ-Data2vec still achieves the best results, with an average WER reduction of ${1.55\%}$ in the three languages es, fr, and it. This reduction is lower than that of PER because UniData2vec concatenates a Transcoder, which is a phoneme-to-word translation model trained on external text. 
Moreover, comparing our DQ-Data2vec in self-supervised and weakly-supervised scenarios, the WER reduction is less than PER.
This discrepancy stems from the label we use is the phonemes of the non-target language (en), this indirect approach limits the model's ability to enhance word-level performance in the target language.

\subsection{Quantization Analysis}
In this section, we conduct a comprehensive analysis of the characteristics of the online K-means quantizer and compare it with previous studies~\cite{xue2023unidata2vec}. To measure the quality of the quantizer in our proposed DQ-Data2vec, we utilize the four metrics introduced in~\cite{hsu2021hubert}:
\begin{itemize}
    \item \textit{Phoneme purity (PP)} measures the purity of the set of associated phones of each codeword.
    \item \textit{Language purity (LP)} similar to PP, measures the purity of the set of associated languages of each codeword.
    \item \textit{Phoneme-normalized mutual information (PNMI)} measures the uncertainty reduction for the underlying
phone when observing the codeword of a frame.
    \item \textit{Language-normalized mutual information (LNMI)}, similar to PNMI, measures the uncertainty reduction for the underlying language when observing the codeword of an utterance.
\end{itemize}
The calculation of the above PP and PNMI requires frame-level phoneme force alignment. We used the Montreal Forced Aligner\footnote{Available at \url{https://mfa-models.readthedocs.io/en/latest/}} (MFA) tool to generate the phoneme alignment on the CommonVoice test set.

\begin{table}[h]
\centering
\footnotesize
\caption{Metrics of the language and phoneme quantizers. SD and DD in the table refer to shallow decoupling and deep decoupling. The four metrics in the table all indicate better performance as they increase in value.}
\label{tab:qt_metrics}
\begin{tabular}{@{}lcccc@{}}
\toprule
\textbf{Model} & \textbf{LP} & \textbf{PP} & \textbf{LNMI} & \textbf{PNMI} \\ \hline\hline
\multicolumn{5}{l}{\textit{\textbf{Offline K-means Clustering}}} \\ \midrule
UniData2vec~\cite{xue2023unidata2vec} & - & 0.37 & - & 0.26 \\ \midrule
\multicolumn{5}{l}{\textit{\textbf{Online K-means Clustering}}} \\ \midrule
DQ-Data2vec (SD) & 0.54 & 0.30 & 0.34 & 0.29 \\ \midrule
DQ-Data2vec (DD) & 0.88 & 0.48 & 0.95 & 0.20 \\ \bottomrule
\end{tabular}
\end{table}

The detailed metrics presented in Table~\ref{tab:qt_metrics} reveal that in the shallow decoupling scenario, the PP of DQ-Data2vec is slightly lower than that of UniData2vec. 
This is due to the larger phoneme count in the dictionary we utilized compared to that of UniData2vec, which does not encompass the phonemes from the sv language. Generally, a higher phoneme count results in a lower PP. However, the metric PNMI can mitigate the influence of the phoneme amount, indicating superior clustering performance of our DQ-Data2vec.
Furthermore, an LNMI of 0.34 illustrates that DQ-Data2vec can adeptly extract language information without depending on language labels. Moreover, in the deep decoupling scenario, both LP and LNMI demonstrate substantial improvement, signifying a significant enhancement of the clustering effect in the language quantizer. However, for the phoneme quantizer, although PP has increased, PNMI has decreased, leading to a conflict. 
Upon closer examination, we find that the confidence of the more probable codewords for a specific phoneme has notably increased, while the confidence of the less probable codewords has significantly decreased. The latter constitutes a much larger proportion, resulting in an overall reduction in PNMI. In essence, the addition of CTC loss causes the phoneme quantizer's clustering to become more concentrated on a few codewords, rendering more codewords ineffective or less efficient. This is primarily because the CTC loss is trained solely by non-target language text labels and does not cover all the phonemes in the dictionary. Nonetheless, for the student branch, reducing effective codewords and highlighting the key codewords means a reduction in the reconstruction difficulty of the mask token. Consequently, the overall auxiliary effects of the phoneme quantizer in the ASR task are enhanced in the deep decoupling scenario.
\begin{figure}[htp]
    \centering
    \subfigure[shallow decoupling]{
        \includegraphics[width=20em]{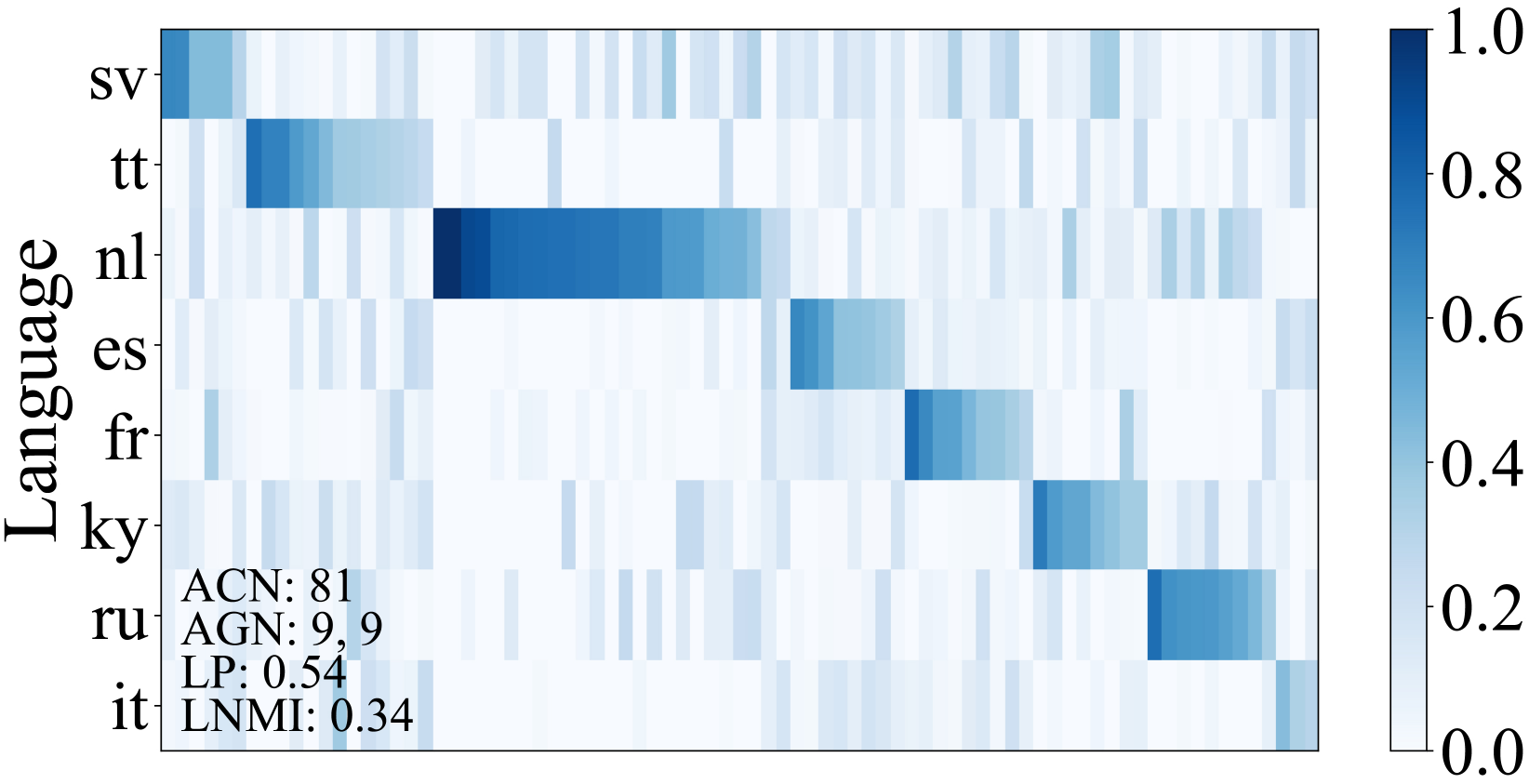}
    }
    \\
    \subfigure[deep decoupling]{
        \includegraphics[width=20em]{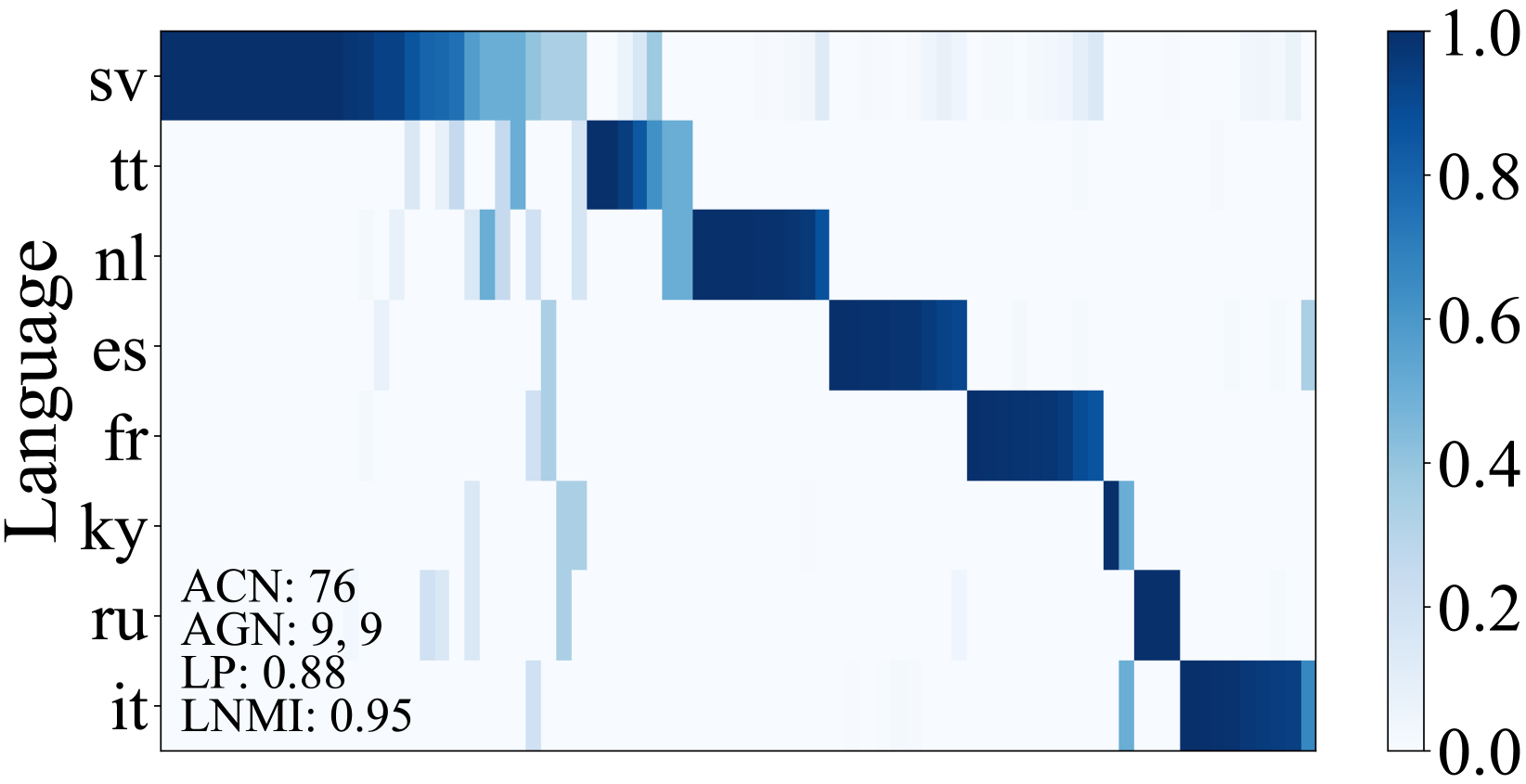}
    }
    \caption{The conditional probability ${P(lang|code)}$ on CommonVoice test set. The y-axis is the language set sorted by the number of occurrences, and the x-axis is the active codewords sorted by the most correlated language. In the figure, ACN denotes the active codeword number, while AGN indicates the active clusters within the two clustering groups. LP signifies language purity, and LNMI refers to language-normalized mutual information.}
    \label{fig:lang_align}
    \vspace{-1em}
\end{figure}

\begin{figure}[htp]
    \centering
    \subfigure[shallow decoupling]{
        \includegraphics[width=20em]{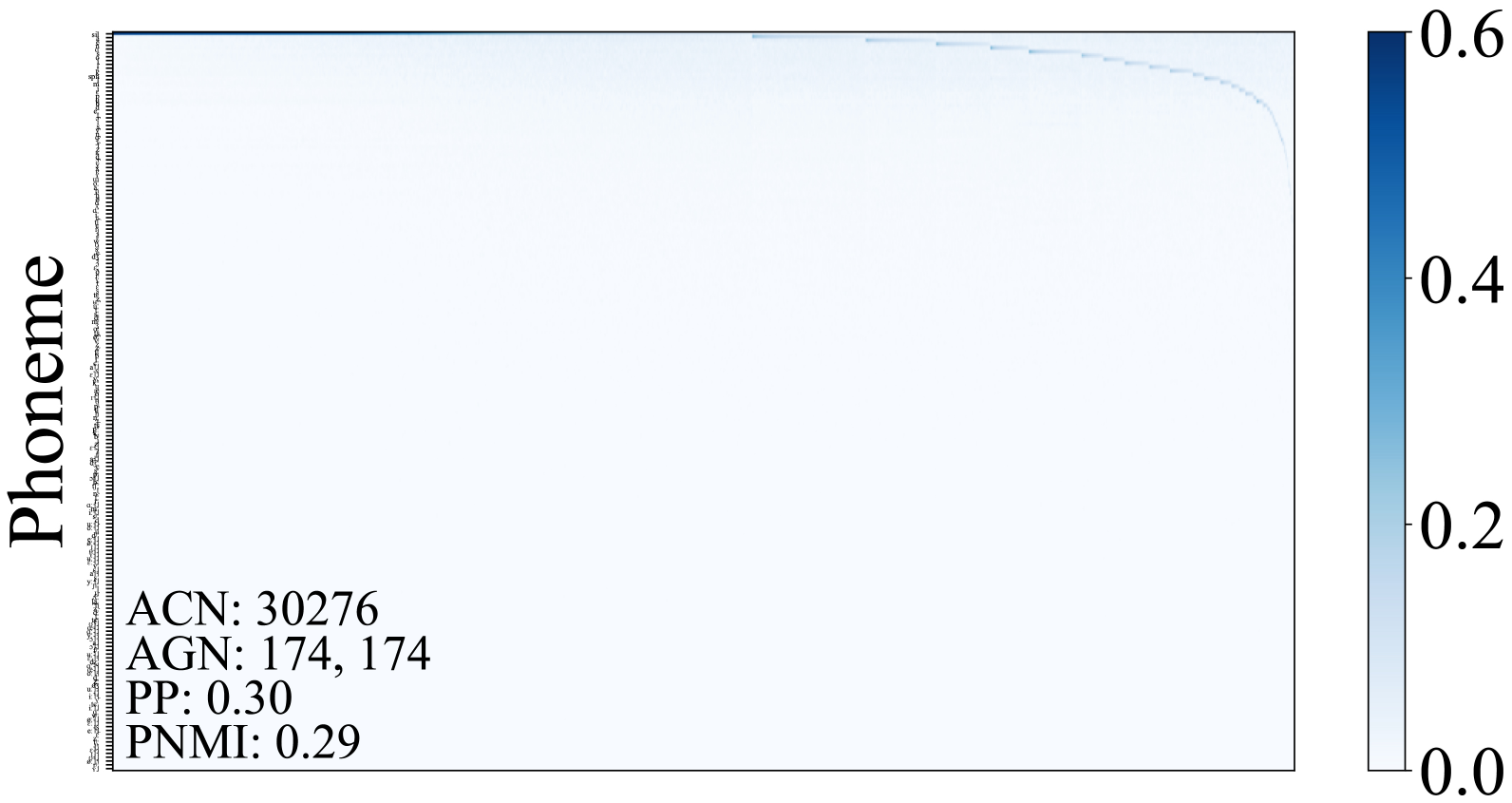}
    }
    \\
    \subfigure[deep decoupling]{
        \includegraphics[width=20em]{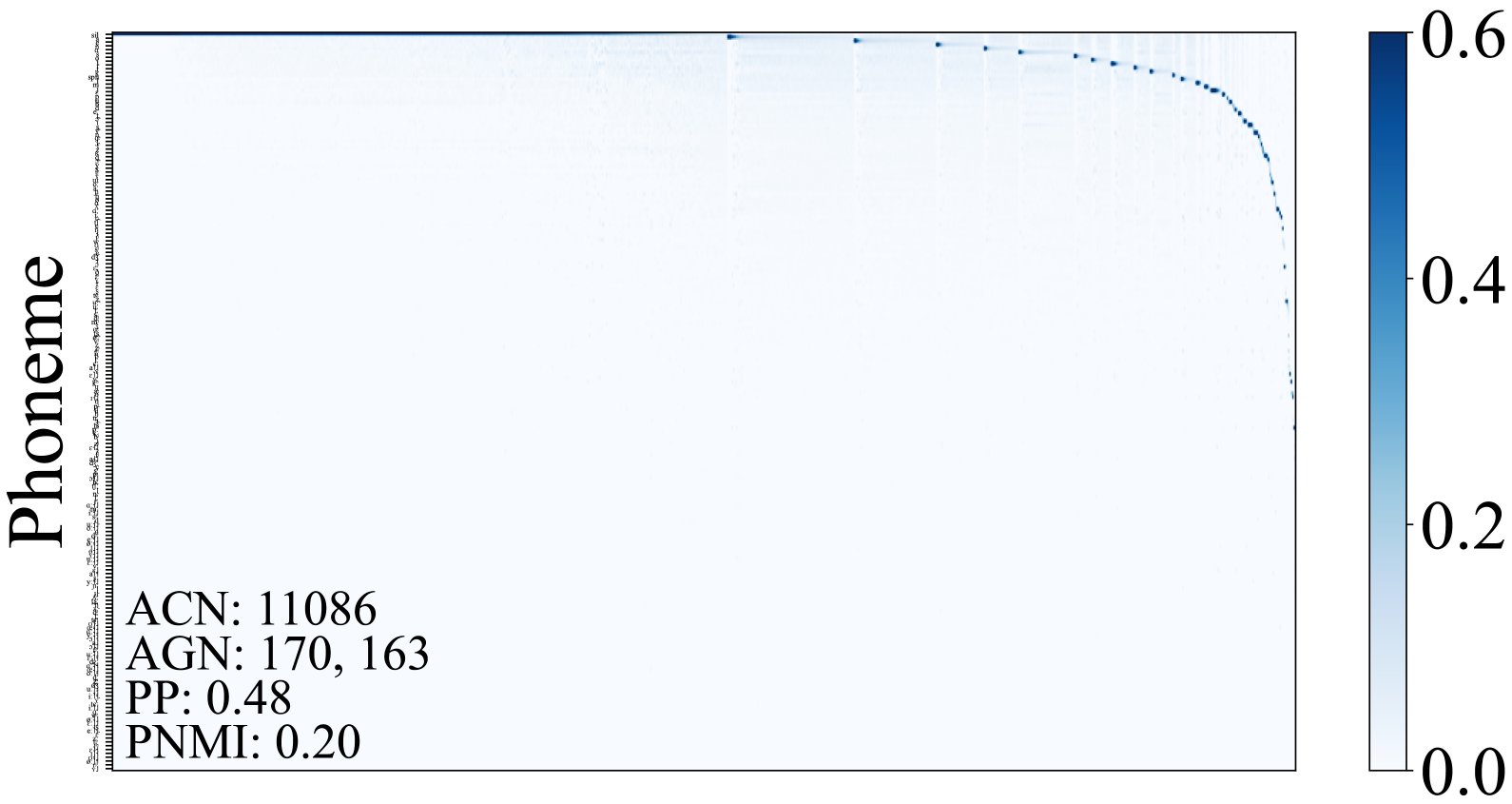}
    }
    \caption{The conditional probability ${P(phone|code)}$ on CommonVoice test set. The y-axis is the phoneme set sorted by the number of occurrences, and the x-axis is the active codewords sorted by the most correlated phoneme. In the figure, ACN denotes the active codeword number, while AGN indicates the active clusters within the two clustering groups. PP signifies phoneme purity, and PNMI refers to phoneme-normalized mutual information.}
    \label{fig:phone_align}
    \vspace{-1em}
\end{figure}

To further illustrate the quality of the learned codebooks, we present the visualizations of the conditional probabilities ${P(lang|code)}$ and ${P(phone|code)}$ aggregated across the CommonVoice test set in Fig.~\ref{fig:lang_align} and Fig.~\ref{fig:phone_align}. In these visualizations, a more concentrated vertical distribution within a language or phoneme indicates higher LP or PP, while a clearer diagonal path typically signifies higher LNMI or PNMI.
As depicted in Fig.~\ref{fig:lang_align}, even in the shallow decoupling scenario, the language quantizer captures distinct alignment paths, providing strong evidence that our DQ-Data2vec can achieve effective language clustering without labels. Moreover, the introduction of CE loss significantly contributes to guiding the quantizer to learn language-specific information, as evidenced by the noticeably clearer paths when language labels are added.
In Fig.~\ref{fig:phone_align}, it is evident that the first row occupies a wide range, indicating that a substantial number of active codewords are allocated to the silencing unit, which is consistent with previous research~\cite{baevski2020wav2vec}. Additionally, audio segments at the beginning and end of the recordings in the CommonVoice dataset contain a significant amount of silence, further exacerbating this phenomenon. Furthermore, both paths in Fig.~\ref{fig:phone_align} are situated in the upper right corner, indicating that many phonemes are not assigned to an adequate number of active codebooks, primarily due to the imbalance of data between languages.
Finally, comparing Fig.~\ref{fig:phone_align} (a) and Fig.~\ref{fig:phone_align} (b), it is evident that the conditional probabilities of active codewords corresponding to a given phoneme exhibit a long-tail effect in each row. In the shallow decoupling scenario, the conditional probabilities show a gentle decline. 
In contrast, in the scenario of deep decoupling, the probability of head codewords increases, while the tail declines more rapidly. This validates our previous assertion that the clustering of the phoneme quantizer becomes more concentrated on a few codewords.
Additionally, as depicted in Fig.~\ref{fig:phone_align}, it is evident that despite the decrease in PNMI of the phoneme quantizer influenced by the weaker tail in the deep decoupling scenario, the alignment paths become more distinct, demonstrating the efficacy of the phoneme quantizer for the ASR task in this scenario. 
Considering that there are often unused codewords in clustering algorithms, aside from the LP/PP and LNMI/PNMI metrics, we also display the active codeword numbers and active cluster numbers in Fig.~\ref{fig:lang_align} and Fig.~\ref{fig:phone_align}. 
It's noteworthy that we divide the feature dimension of acoustic features into two groups, with each group clustered independently, and the number of clusters in each group being either 9 for languages or 174 for phonemes. Consequently, the maximum actual used codeword numbers for the language quantizer and the phoneme quantizer are ${9\times9 = 81}$ and ${174\times174 = 30276}$ respectively.
When the active codeword numbers do not reach the maximum, it is usually not equal to the product of two active cluster numbers, because the two clusters are counted independently.
As seen in Fig.~\ref{fig:lang_align} (a) and Fig.~\ref{fig:phone_align} (a), in the shallow decoupling scenario, all codewords are activated. This is mainly because, as indicated in Equation~\ref{eq:loss_ctr}, we introduce contrastive loss during the training of quantizers, which compels the assignment of as many codewords as possible to the one phoneme or language. However, as shown in Fig.~\ref{fig:lang_align} (b) and Fig.~\ref{fig:phone_align} (b), with the supervised losses introduced into the deep decoupling scenario, the numbers of active codewords decline. This is reasonable as the certainty of active codewords is higher. In summary, while a great number of active codewords does not always guarantee superior clustering, too few quantities indicate clustering collapse. Evidently, our clustering results do not fall into the latter scenario. Moreover, in both the shallow decoupling and deep decoupling scenarios, most of our clusters are active, which proves the feasibility of specifying the number of clusters as the language or phoneme types.

In addition, considering that in Fig.~\ref{fig:phone_align} (a), the alignment paths do not appear clear enough, we separate the multilingual test set by language and compute their individual PP and PNMI metrics. The results in Table \ref{tab:single_lang_metrics} show that individual PP and PNMI significantly surpass their corresponding metrics in the multilingual mix test sets. This disparity is primarily attributable to the unbalanced distribution of phonemes. Specifically, our analysis reveals that the number of phonemes within the language-individual test set ranges from 29 to 90, whereas the multilingual test sets contain a total of 174 phonemes, indicating that many phonemes occur in only one language. 
Nevertheless, whether in the language-individual or multilingual test sets, an occurrence probability of silence ('sil') of approximately ${33\%}$.
Consequently, in a language-individual test set, the remaining 28-89 phonemes share approximately ${67\%}$ of the rest probability, while in the multilingual test set, the 173 phonemes share this ${67\%}$ probability. Clearly, in the multilingual test set, the observation probability of each phoneme is much lower than that in the language-individual test set, which results in poor PNMI and PP. 
When examining the clustering performance of individual languages, our phoneme quantizer's PNMI significantly increases, ranging from 0.49 to 0.62. For reference, HuBERT's PNMI in English is ${0.704}$~\cite{hsu2021hubert}. Although our results are still lower than HuBERT, they also underscore the effectiveness of the phoneme clustering in our DQ-Data2vec.

\begin{table}[h]
\centering
\caption{Phoneme quantizer
metrics of shallow decoupling in language-individual tests.}
\label{tab:single_lang_metrics}
\resizebox{\linewidth}{!}{
\begin{tabular}{@{}lccccccccl@{}}
\toprule
\textbf{Metric} & \textbf{es} & \textbf{fr} & \textbf{it} & \textbf{ky} & \textbf{tt} & \textbf{nl} & \textbf{ru} & \textbf{sv} & \textbf{Avg} \\ \hline\hline
PP & 0.59 & 0.56 & 0.58 & 0.55 & 0.47 & 0.57 & 0.48 & 0.60 & 0.55 \\ \midrule
PNMI & 0.56 & 0.58 & 0.60 & 0.56 & 0.49 & 0.60 & 0.59 & 0.62 & 0.58 \\ \bottomrule
\end{tabular}}
\end{table}

\subsection{Comparison Experiments for Decoupling Quantization}
In this section, we present the comparison experiments of the diverse designs for our decoupling quantization, showcasing the rationale behind the design in our DQ-Data2vec and offering a reference for similar SSL model designs.
As depicted in Table~\ref{tab:compar}, we initially examine the influence of quantizer types. Gumbel-softmax quantization has been demonstrated as an effective approach for generating discrete speech units~\cite{baevski2020vq,baevski2020wav2vec}. However, when comparing ${S1}$ and ${B2}$, we observe that their PNMIs are similar, but the PER using Gumbel-softmax quantization is higher than with online K-means. In Gumbel-softmax quantization, the absence of Euclidean distance constraints between the layer outputs of the teacher branch and the quantizers' codewords results in the inability to ensure that the feature dimension distribution of the latter is similar to that of the former.
Consequently, the codewords in Gumbel-softmax quantization cannot fully capture the feature-dimension details of language or phoneme information. In essence, the quantization vectors of Gumbel softmax concentrate more on the index of the codewords rather than the content of the feature dimensions of the codewords. These contents are not crucial for performance in wav2vec 2.0~\cite{baevski2020wav2vec} and vq-wav2vec~\cite{baevski2020vq}, but they are significant in our DQ-Data2vec.
Next, we investigate the group number of the temporal convolution, which is utilized to simulate product quantization as mentioned in Section~\ref{sec:Method}.B. In the construction of the standard baseline ${B3}$, we designate the group number as 2 to divide the feature dimensions of the teacher's layer results into 2 segments, each undergoing independent clustering and then concatenated. Upon comparison of ${B3}$ with ${S2}$, it becomes evident that when group=1, both the clustering and ASR performance experience a significant decrease. 
This underscores the importance of product quantization.
Then, as discussed in Section~\ref{sec:Method}.B, we compare the effects of different normalizations on frame-level and utterance-level quantization in ${S3}$ and ${S4}$. When L2 normalization is applied for frame-level quantization, the pre-training phase experiences significant instability and a loss explosion at approximately 350k updates, leading to the inability to complete the training. On the other hand, when instance normalization is utilized in utterance-level quantization, pre-training proceeds as normal, but the LNMI decreases notably to 0.09 from 0.42, subsequently resulting in a significant increase in PER. This suggests that the language quantizer fails to capture sufficient language information with the instance normalization. These two experiments illustrate that the choice of suitable normalization for quantization at various levels is a crucial factor that impacts clustering.
Finally, we demonstrate the impact of the trainable parameters in the gray block depicted in Fig.~\ref{fig:system_framework} (b). In shallow decoupling ${S5}$, we initially introduce a Transformer layer, resulting in rapid collapse shortly after the beginning of pre-training. This collapse is specifically reflected in the fact that the codebook index sequence output by the quantizer is similar regardless of the input audio. Upon replacing the Transformer layer with CNN in ${S6}$, the collapse occurs at a slower rate, yet the phenomenon remains consistent. This indicates that the addition of excessive trainable parameters in the teacher branch can cause the quantizer to learn pseudo-solutions, ultimately leading to collapse. The Transformer layer possesses a more robust context learning capability than CNN, rendering it easier to generate pseudo-solutions that fulfill the self-supervised loss.
However, in the case of deep decoupling ${S7}$, the result is contrary. With the presence of supervised losses, the collapse problem is no longer present. In this scenario, the inclusion of CNN can significantly enhance the performance of clustering and ASR.
\begin{table}[htp]
\centering
\footnotesize
\caption{Comparison and ablation experiments. The gray area represents the deep decoupling scenario, and the rest is shallow decoupling.}
\label{tab:compar}
\resizebox{.99\hsize}{!}{$
\begin{tabular}{@{}llccccc@{}}
\toprule
\textbf{ID} & \textbf{Model} & \textbf{LQT} & \textbf{PQT} & \multicolumn{1}{l}{\textbf{PER}${\downarrow}$} & \textbf{LNMI}${\uparrow}$ & \textbf{PNMI}${\uparrow}$ \\ \hline\hline
\multicolumn{7}{l}{\textit{\textbf{Standard Baseline}}} \\ \midrule
B1 & DQ-Data2vec w/o PQT & \cmark & \xmark & 7.31 & 0.42 & \xmark \\ \midrule
B2 & DQ-Data2vec w/o LQT & \xmark & \cmark & 7.63 & \xmark & 0.28 \\ \midrule
B3 & DQ-Data2vec & \cmark & \cmark & 7.23 & 0.34 & 0.29 \\ \midrule
\rowcolor[HTML]{EFEFEF}
B4 & DQ-Data2vec (DD) & \cmark & \cmark & 6.93 & 0.95 & 0.20 \\ \midrule
\multicolumn{7}{l}{\textit{\textbf{Comparison of Quantizer Type}}} \\ \midrule
S1 & Gumbel Softmax & \xmark & \cmark & 7.85 & \xmark & 0.30 \\ \midrule
\multicolumn{7}{l}{\textit{\textbf{Comparison of Group Number}}} \\ \midrule
S2 & Group=1 & \cmark & \cmark & 7.40 & 0.13 & 0.12 \\ \midrule
\multicolumn{7}{l}{\textit{\textbf{Comparison of Norm}}} \\ \midrule
S3 & Use L2Norm in PQT & \xmark & \cmark & \multicolumn{3}{c}{loss explosion} \\ \midrule
S4 & Use InsNorm in LQT & \cmark & \xmark & 7.68 & 0.09 & \xmark \\ \midrule
\multicolumn{7}{l}{\textit{\textbf{Trainable Parameters in Quantizer}}} \\ \midrule
S5 & Add Transformer & \xmark & \cmark & \multicolumn{3}{c}{quick collapse} \\ \midrule
S6 & Add CNN & \xmark & \cmark & \multicolumn{3}{c}{slow collapse} \\ \midrule
\rowcolor[HTML]{EFEFEF}
S7 & Del CNN (DD) & \cmark & \cmark & 7.13 & 0.63 & 0.32 \\ \bottomrule
\end{tabular}$}
\end{table}

\section{Conclusions}
\label{sec:Conclusions}
In this research, we propose DQ-Data2vec, a decoupling quantization SSL approach for multilingual ASR. This method is specifically designed to acquire phonemes and language knowledge to enhance the ASR task in multilingual SSL.
The approach incorporates a teacher-student-based backbone and two improved online K-means vector quantizers. In the shallow decoupling scenario, two K-means quantizers extract discrete language and phoneme-related speech representations from different layers' results of the teacher, with a constraint on the quantizer object by specifying the codebook numbers and pooling. Furthermore, in the deep decoupling scenario, we investigate the inclusion of language labels and non-target language text labels to establish a more explicit relationship between the quantization vector and phonemes or language. Our experiments on the multilingual CommonVoice dataset demonstrate the effectiveness of the proposed model. In both shallow and deep decoupling scenarios, 
DQ-Data2vec achieves a relative reduction of ${9.51\%}$/${18.09\%}$ in PER and ${11.58\%}$/${1.55\%}$ in WER over data2vec and state-of-the-art (SOTA) UniData2vec baselines, respectively. Looking ahead, we aim to explore the use of codebook discrete units in decoupling quantization for large language models to further expand the applicability of our DQ-Data2vec.

\bibliographystyle{Transactions-Bibliography/IEEEtran}
\bibliography{Ref.bib}
%




\end{document}